\documentclass[english]{jssst_ppl}

\usepackage{cite}
\usepackage{amsmath,amssymb,amsfonts}
\usepackage{algorithmic}
\usepackage{graphicx}
\usepackage{textcomp}
\usepackage{xcolor}
\def\BibTeX{{\rm B\kern-.05em{\sc i\kern-.025em b}\kern-.08em
    T\kern-.1667em\lower.7ex\hbox{E}\kern-.125emX}}
\usepackage{latexsym}
\newtheorem{definition}{Definition}
\usepackage{booktabs} 
\usepackage{multirow}
\usepackage{diagbox}
\usepackage[braket]{qcircuit} 
\usepackage{threeparttable} 
\usepackage{stfloats} 
\usepackage{enumitem} 
\usepackage{fancyheadings} 

\usepackage{makecell}
\usepackage{times}
\usepackage{graphicx}
\usepackage{epsf}
\usepackage{verbatim}
\usepackage{url}
\usepackage{color}
\usepackage{alltt}

\newcommand{\Comment}[1]{}

\newcommand{\SmallSpace}{\vspace*{-1.4ex}}

\usepackage{xcolor}

\usepackage{braket} 
\usepackage{listings}
\usepackage{color}
\usepackage{float} 
\usepackage{subfigure}
\lstdefinelanguage{QSharp}
{morekeywords={operation, Int, if, body, Unit, let, for, in, Controlled, adjoint, Adj, Ctl, is, Qubit},
sensitive=true,
morecomment=[l]{//},
}
\def\BibTeX{{\rm B\kern-.05em{\sc i\kern-.025em b}\kern-.08em
    T\kern-.1667em\lower.7ex\hbox{E}\kern-.125emX}}

\title{Static Entanglement Analysis of Quantum Programs
}

\author{Shangzhou Xia, Jianjun Zhao}
\inst{%
Kyushu University\\
\texttt{\{xia.shangzhou.218@s,zhao@ait\}.kyushu-u.ac.jp}
}

\begin{document}
\maketitle

\begin{abstract}
Quantum entanglement plays a crucial role in quantum computing. Entangling information has important implications for understanding the behavior of quantum programs and avoiding entanglement-induced errors. Entanglement analysis is a static code analysis technique that determines which qubit may entangle with another qubit and establishes an entanglement graph to represent the whole picture of interactions between entangled qubits. This paper presents the first static entanglement analysis method for quantum programs developed in the practical quantum programming language Q\#. Our method first constructs an interprocedural control flow graph (ICFG) for a Q\# program and then calculates the entanglement information not only within each module but also between modules of the program. The analysis results can help improve the reliability and security of quantum programs.
\end{abstract}

\section{Introduction}

In recent years, more and more quantum programming languages and environments, such as ~\cite{svore2018q,cirq2018google}, have been developed to support programming quantum computers. Because quantum programming requires exploiting unique quantum properties such as superposition and entanglement, it is considered more challenging than classical programming, which makes understanding the behavior of quantum programs very difficult. Therefore, there is an urgent need to develop methods and tools to efficiently and automatically support the analysis of quantum programs.

Quantum entanglement plays a crucial role in quantum computation. Entanglement relations between qubits in a quantum program have important implications for understanding the behavior of the quantum program and avoiding entanglement-induced errors in the program. 
%
%
During the execution of a quantum program, it is inevitable that some qubits need to be measured. However, performing measurements without considering the entanglement information between qubits may lead to unexpected state corruption of some qubits in the program, resulting in program errors and information loss. In addition, due to the no-cloning principle of quantum computing, ancilla qubits are often used to assist the operation during program execution. After completing the auxiliary operation, it is necessary to ensure that the ancilla qubits are no longer entangled with other qubits in the program. 

Several entanglement analysis methods~\cite{perdrix2007quantum,perdrix2008quantum,honda2015analysis,javadiabhari2015scaffcc,prost2009reasoning} have been proposed, but no entanglement analysis method has been available for the practical quantum programming language Q\# until now. In this paper, we present \textit{JiuChan}, an efficient static entanglement analysis method for Q\# programs. Our method first constructs an interprocedural control flow graph (ICFG) for a Q\# program and then calculates the entanglement information within each module and between modules of the program. Our method establishes an entanglement graph to represent the whole picture of interactions between entangled qubits in the program. The analysis results can help improve the reliability and security of quantum programs.

Our main contribution in this paper is a novel interprocedural entanglement analysis algorithm for quantum programs in Q\#. This algorithm, to the best of our knowledge, is the first entanglement analysis algorithm for quantum programs written in a practical quantum programming language such as Q\#.
  

The rest of the paper is organized as follows. Section~\ref{sec:quantum-computation} introduces some basic concepts of quantum computation and Q\#. Section~\ref{sec:example} presents an example that illustrates how the algorithm of entanglement analysis works. Section~\ref{sec:entanglement analysis} presents the analysis framework and algorithm. 
Section~\ref{sec:implementation} discusses some implementation issues. 
Related work is discussed in Section~\ref{sec:related-work}, and the conclusion is given in Section~\ref{sec:conclusion}.

\section{Backgroud Information}
\label{sec:quantum-computation}
We briefly introduce some basics of quantum computing~\cite{nielsen2002quantum}.

\subsection{Basic Concepts of Quantum Computation}

\subsubsection{Quantum Bit}
\label{subsec:qubit}

A classical bit is a binary unit of information used in classical computation. It can take two possible values, 0 or 1. A quantum bit (or \textit{qubit}) is different from the classical bit in that its state is theoretically represented by a linear combination of two bases in the quantum state space (represented by a column vector of length 2). We can define two qubits $|0\rangle$ and $|1\rangle$, which can be described as

\begin{center}
$|0\rangle = \begin{bmatrix} 1 \\ 0  \end{bmatrix}$ and $|1\rangle = \begin{bmatrix}0 \\1 \end{bmatrix}$
\end{center}

\vspace*{0mm}
\noindent
Qubits $|0\rangle$ and $|1\rangle$ are the computational basis state of the qubit. In other words, they are a set of the basis of quantum state space.

Any qubit $|e\rangle$ can be expressed as a linear combination of two bases:

\begin{center}
$|e\rangle = \alpha|0\rangle + \beta|1\rangle$
\end{center}

\vspace*{0mm}
\noindent
where $\alpha$ and $\beta$ are complex numbers, and $|\alpha|^2+|\beta|^2 = 1$. This restriction is also called {\it normalization conditions}.

\subsubsection{Quantum Gate and Circuit}
\label{subsec:q-gate}

Just as a logic gate in a digital circuit can modify the state of a bit, a quantum gate can change the state of a qubit. A quantum gate can have only one input and one output (transition of a single quantum state), or it can have multiple inputs and multiple outputs (transition of multiple quantum states). The number of inputs and outputs should be equal because the operators need to be reversible, which means no information can be lost in quantum computing. 

\vspace*{0mm}
\noindent
\textit{NOT Gate.}\hspace*{0.6mm} The NOT gate works on a single qubit, which can exchange the coefficients of two basis vectors: $NOT(\alpha |0\rangle + \beta |1\rangle) = \alpha |1\rangle + \beta |0\rangle$. The quantum NOT gate is an extension of the NOT gate in classical digital circuits. 

A single input-output quantum gate can be represented by a $2\ \times\ 2$ matrix. The state of a quantum state after passing through the quantum gate is determined by the value of the quantum state vector left multiplied by the quantum gate matrix. The quantum gate matrix corresponding to the NOT gate is

\begin{center}
$X = \begin{bmatrix}0&1\\1&0\end{bmatrix}$. 
\end{center}

\noindent
Therefore, the result of a qubit passing a NOT gate can be denoted as 

\begin{center}
$X \begin{bmatrix}\alpha \\ \beta \end{bmatrix} = \begin{bmatrix}0&1\\1&0\end{bmatrix} \begin{bmatrix}\alpha \\ \beta \end{bmatrix} = \begin{bmatrix}\beta \\ \alpha \end{bmatrix}$.
\end{center}

\vspace*{0mm}
\noindent
\textit{Hadamard Gate.}\hspace*{0.6mm}The Hadamard gate also works on a single qubit, which can decompose existing quantum states according to its coefficients as: 

\begin{center}
$H(\alpha |0\rangle + \beta |1\rangle) = \frac{\alpha + \beta}{\sqrt{2}}|0\rangle + \frac{\alpha - \beta}{\sqrt{2}}|1\rangle$. 
\end{center}

\vspace*{0mm}
\noindent
This can be represented by a matrix: 

\vspace*{0mm}
\begin{center}
$H = \frac{\sqrt{2}}{2}
\begin{bmatrix}1&1\\1&-1\\\end{bmatrix}$. 
\end{center}

\vspace*{0mm}
\noindent
Although the Hadamard gate is not directly related to the AND and OR gates in classical digital circuits, it has important applications in many quantum computing algorithms. 

There are two properties in qubit: magnitude and phase. Thus, quantum gates can be divided into magnitude gates and non-magnitude gates according to the function area. The above-mentioned NOT gate and Hadamard gate belong to magnitude gates because they can change the magnitude of the target qubit and, thus, the probability of the state output. In addition, the magnitude gate can also change the control relationship. In contrast, a non-magnitude gate can only change the phase of the qubit, such as a T gate, or Z gate. Phase information cannot be reflected in the state probability; therefore, some algorithms will use the combination of Hadamard gates to handle the phase information.

\subsection{Quantum Entanglement}
\label{subsec:q-entanglement}
Quantum systems may exhibit {\it entanglement}~\cite{einstein1935can,schrodinger1935discussion}, a quantum mechanical phenomenon. A state is considered entangled if it cannot be broken down into more basic parts. The existence of entanglement relations makes mutually independent qubit systems connected, thus enabling the information interaction of different qubit systems. At the same time, the existence of entanglement relations among systems makes it impossible to be considered separable systems. When a measurement is made on some qubits in the system, it affects the state of other qubits. 

\begin{definition}
For a state $| \varphi \rangle$ of a set $S$ of qubits and a partition $(A, B)$ of $S$, $(A, B)$ is \textit{entangled} iff there does not exist two states $| \varphi_{A} \rangle$ and $| \varphi_{B} \rangle$ of the respective parts $A$ and $B$ such that $| \varphi \rangle$ = $| \varphi_{A} \rangle$ $\displaystyle \otimes$ $| \varphi_{B} \rangle$. $(A, B)$ is \textit{separable} iff $(A, B)$ is not entangled.
\end{definition}

For example, a GHZ state $\frac{\sqrt{2}}{2}(|000 \rangle +|111 \rangle )$ is entangled. When the result of observing the first qubit is 0, the other qubits must all be 0. Therefore, the measurement of some qubits in the absence of system entanglement information leads to the destruction of the system state in which it is located, triggering bugs and information loss in the program.

According to~\cite{perdrix2007quantum}, for a state $| \varphi \rangle$ of a set $S$ of qubits and a partition $(A, B, C)$ of $S$. The entanglement relation has the following three properties.

\begin{itemize}
\item \textbf{Transitive}: If $(A, B)$ is entangled and $(B, C)$ is entangled, then $(A, C)$ is entangled.
\item \textbf{Symmetric}: If $(A, B)$ is entangled, then $(B, A)$ is entangled 
\item \textbf{Eliminable}: If $(A, B)$ is separable and for an operation $U$ that $U(| \varphi_{A} \rangle \displaystyle \otimes | \varphi_{B} \rangle)$ is entangled, then there must exist operation $V$ that $VU(| \varphi_{A} \rangle \displaystyle \otimes | \varphi_{B} \rangle)$ is separable.
\end{itemize}


\subsection{Uncomputation Mechanism}
\label{subsec:uncomputation}

Since quantum operations are unitary operations, there must be a corresponding inverse operation for each operation step.
Uncomputation is the mechanism by which an operation cancels out when the operation and the corresponding inverse operation occur simultaneously. The description using the matrix is that the multiplication of two operation matrices results in an identity matrix. Due to the no-cloning nature of qubits, the information in a qubit can only be used but cannot be copied. Therefore, each operation step modifies the information in the original qubit, and the uncomputation mechanism is used to restore it after the execution of the operation to ensure that the original information remains unchanged. In addition to this, for the intermediate results of operations in quantum programs, a new qubit is needed to save the result temporarily, that is, the \textit{ancilla qubit}, and the ancilla qubit needs to be restored after the operation to ensure that the ancilla qubits does not affect the program. This process also uses the uncomputation mechanism. 
For some common gate operations, their corresponding inverse operations are shown in Table~\ref{table:inverse-operation}.

\begin{table}[h]
\caption{Some quantum gates and their corresponding inverse gates.}
\label{table:inverse-operation}
\begin{center}
\scriptsize
\renewcommand\arraystretch{1.2}
\begin{tabular} {p{2.5cm}|p{3cm}}
\hline 
\textbf{Operation} & \textbf{Inverse Operation
} \\\hline 
\hline
Hadamard & Hadamard \\\hline 
NOT & NOT \\\hline 
Phase(a) & Phase(-a) \\\hline 
CNOT(a, b) & CNOT(a, b) \\\hline 
\end{tabular}
\end{center}
\end{table}

The Q\# language allows user-defined operations, such as the \texttt{GHZ} operation in the example in Figure~\ref{fig:example}. The syntactic structure of a user-defined operation is shown in Figure~\ref{fig:template}. For the \texttt{Name} operation, Q\# will have a corresponding adjoint \texttt{Name} operation. When the \texttt{Name} operation and the adjoint \texttt{Name} operation act on the same qubit in succession, the uncomputation mechanism is implemented to cancel each other out. It should be noted that since the non-magnitude operations do not change the magnitude of the qubit, using non-magnitude operations on the control bit qubit does not change the control relationship. Therefore, between operation pairs that satisfy the uncomputation mechanism, the uncomputation mechanism is still satisfied even if the non-magnitude operation is used for the control bit qubit.

\begin{figure}[htbp]
{\scriptsize
\begin{alltt}
\textcolor{blue}{operation} Name ( Q : \textcolor{blue}{Qubit} ) : \textcolor{blue}{Unit is Adj+Ctl} \{

     \textcolor{blue}{body}(...) \{
       //operation
     \}

     \textcolor{blue}{adjoint}(...) \{
       //inverse operation
     \}

     \textcolor{blue}{Controlled}(cs, ...) \{
       //controlled operation
     \}
 \}
\end{alltt}
  \caption{The template of Q\# operation.}
  \label{fig:template}
}
\end{figure}


  


Due to the existence of the uncomputation mechanism, the entanglement relationship in the program also has the possibility of elimination. Therefore, the uncomputation mechanism must be considered to improve the accuracy of program entanglement analysis.

\section{Example}
\label{sec:example}

We next present an example to illustrate how our entanglement analysis works. 
Figure~\ref{fig:example} is a quantum program written in Q\#. The main body of the program is composed of two operations \texttt{Entangle\_test} and \texttt{GHZ}. Since researchers may modify and reorganize the existing algorithm in the process of developing the algorithm, we construct the structure of calling the GHZ algorithm in the user development program. Also, the program allows the user to input the parameter \texttt{a} as a control for whether to call the \texttt{GHZ} algorithm or not. \texttt{EntryPoint} (line 13) tells the Q\# compiler where to begin executing the program.

\begin{figure}[th]
{\scriptsize
\setlength{\abovecaptionskip}{-0.2cm}
\begin{alltt}
 1    \textcolor{blue}{namespace} NamespaceQFT \{
 2        \textcolor{blue}{open} Microsoft.Quantum.Intrinsic;
 3        \textcolor{blue}{open} Microsoft.Quantum.Diagnostics;
 4        \textcolor{blue}{open} Microsoft.Quantum.Math;
 5        \textcolor{blue}{open} Microsoft.Quantum.Arrays;
 6
 7        \textcolor{blue}{operation} GHZ(target:\textcolor{blue}{Qubit[])}: \textcolor{blue}{Unit} \{
 8            H(target[0]);
 9            \textcolor{blue}{Controlled} X(target[0], target[1]);
10            \textcolor{blue}{Controlled} X(target[1], target[2]);
11        \}
12
13        @EntryPoint()
14        \textcolor{blue}{operation} Entangle_test(a:\textcolor{blue}{Int}) : \textcolor{blue}{Unit} \{
15            \textcolor{red}{use} qs=\textcolor{blue}{Qubit}[4];
16
17            H(qs[0]);
18            X(qs[1]);
19            H(qs[3]);
20            \textcolor{blue}{Controlled} R1([qs[0]], (PI()/2.0, qs[2]));
21            \textcolor{blue}{Controlled} X(qs[0], qs[2]);
22            if a==1:
23                GHZ([qs[0],qs[1],qs[2]]);
24                \textcolor{blue}{Controlled} R1([qs[1]], (PI()/4.0,qs[3]));
25            \textcolor{blue}{Controlled} X(qs[1], qs[0]);
26            if a==1:
27                \textcolor{blue}{Controlled} R1([qs[1]], (-PI()/4.0, qs[3]));
28                \textcolor{blue}{Controlled} X(qs[1], qs[0]);
29            H(qs[3])
30        \}
31    \}
\end{alltt}
  \caption{An example Q\# program.}
  \label{fig:example}
}
\end{figure}

Based on the three entanglement properties introduced in Section~\ref{sec:quantum-computation}, we represent the entanglement relation using an entanglement graph. In the entanglement graph, nodes represent qubits in a superposition state, and edges represent entanglement relations. Two nodes in an entanglement graph are entangled if they are connected. Therefore, we can modify the entanglement graph step by step according to the interprocedural control flow graph (ICFG). The entanglement relation of the whole program will be generated automatically at the end of the ICFG-based analysis.

First, our analysis algorithm constructs the corresponding \textit{control flow graph} (CFG for short) for each \texttt{operation} module in the program. Due to the nature of quantum operations, we transform the statements in Q\# into the following structure, which we call \textit{operation line}.

\vspace*{0mm}
\begin{center}
\texttt{(Functor, Operation, Control, Target)}
\end{center}
\vspace*{0mm}

\noindent


\noindent
Based on the call relationship, we generate the corresponding \textit{interprocedural control flow graph} (ICFG for short). Then, we classify the state of a qubit into the \textit{classical state} (denoted by 0-state and 1-state) and the \textit{quantum state (superposition state)} (denoted by $\mathcal{Q}$-state) according to whether the qubit state is in the superposition state or not. At the same time, due to the uncomputation mechanism, we create a stack data structure for the $\mathcal{Q}$ to record the operations. The state system and state transition rules for qubit will be described in detail in Section IV.

For the Q\# code in Figure~\ref{fig:example}, the program starts executing from the \texttt{Entrypoint()} statement and creates four qubits with 0-state ($\ket{0}$ by default) at the time of the \texttt{use} statement (line-0). When the line-1 operation is received, since \texttt{qs[0]} is in 0-state, the transition rule 0 is executed. The operation satisfies the requirement to change from 0-state to $\mathcal{Q}$-state, so we change the state of \texttt{qs[0]} to $\mathcal{Q}$-state and create the corresponding stack and node in entanglement graph. For the line-2 and line-3 operations, since the target qubit of the operation is not in the $\mathcal{Q}$-state, it is processed similarly to the first operation line by using the transition rule 0 or 1. For the line-4 operation, the target qubit is in 0-state, and the operation is a non-magnitude operation, so it can be ignored. For the line-5 operation, since it satisfies the switching condition of transition rule 0 and also meets the condition of entanglement generation, we convert the state of \texttt{qs[2]} to $\mathcal{Q}$-state and create the corresponding stack and node. Since there is an entanglement relationship, we add the line-5 operation to Entangle and create an edge between them.

When executing the \verb+GHZ+ statement (line-6), we can use the result of the \verb+GHZ+ transformation and the entanglement graph. Since three inputs are required when calling the \texttt{GHZ} function, we create the inputs \verb+(a,b,c)+ with the states ($\mathcal{Q}$, $\mathcal{Q}$, $\mathcal{Q}$) for the three qubits. The same processing is used for the GHZ internal operations, which generate the corresponding stack and graph.

When \texttt{GHZ} is called, an alias relationship between the input qubits (\verb+qs[0]+, \verb+qs[1]+and \verb+qs[2]+) and the \verb+(a,b,c)+ is created:
\vspace*{0mm}
\noindent
\begin{center}
\{ \verb+qs[0]+ $\leftrightarrow$ \verb+a+, \verb+qs[1]+ $\leftrightarrow$ \verb+b+, \verb+qs[2]+ $\leftrightarrow$ \verb+c+ \}. 
\end{center}
\vspace*{0mm}

\begin{figure}[h]
\centerline{\includegraphics[width=0.8\linewidth]{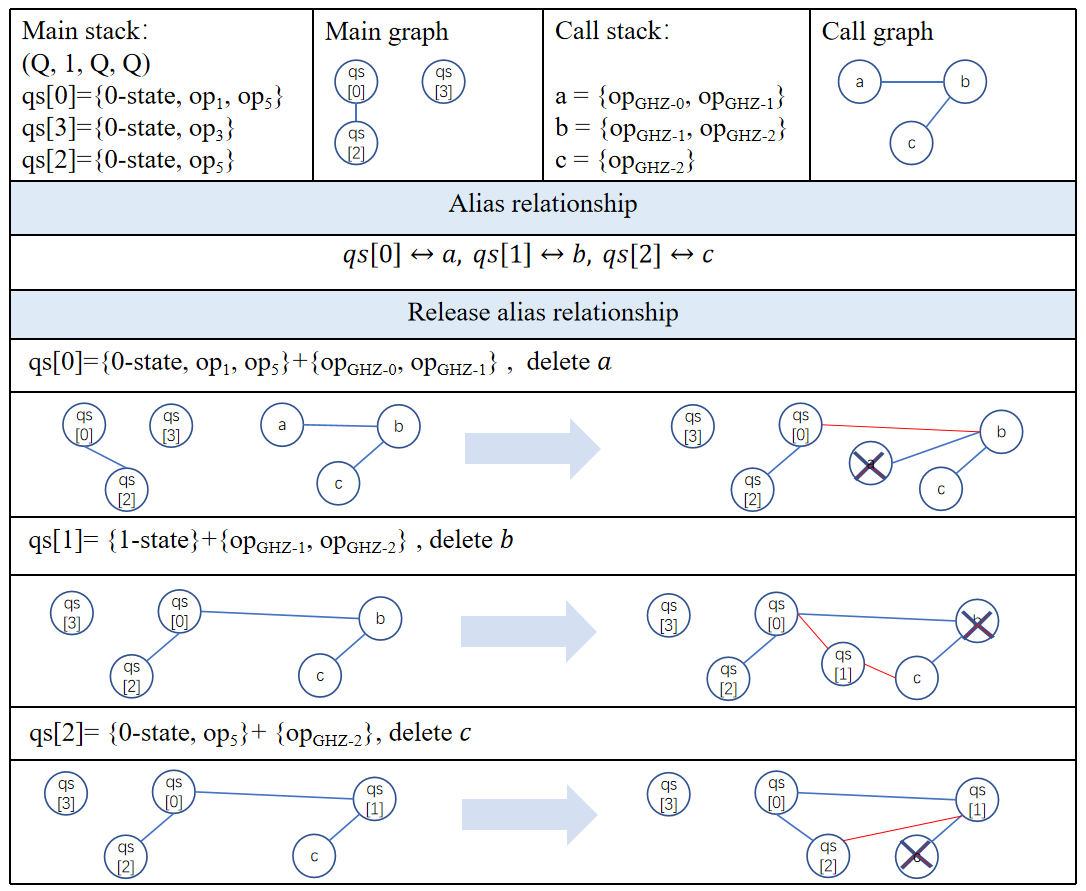}}
\caption{The interprocedural analysis based on the ICFGs for calling the \texttt{GHZ} operation.}
\label{fig:example inter}
\end{figure}

Then the aliasing relation is lifted in turn, e.g., for \texttt{qs[0]}. Now the stack operation of \verb+a+ is passed to \verb+qs[0]+. In the process of stack passing, as in the judgment of entering the stack, it is necessary to detect whether the top of the stack of \verb+qs[0]+ is the inverse operation of the bottom of the stack of \verb+a+. It is necessary to determine whether the state of \texttt{qs[0]} has changed in each pass, as this will affect which transition rule is chosen.  Also, we connect the point connected to \verb+a+ in the entanglement graph to the node of \verb+qs[0]+ and delete the node of \verb+a+, as shown in Figure~\ref{fig:example inter}.

\begin{figure*}[htbp]
\centerline{\includegraphics[width=0.9\linewidth]{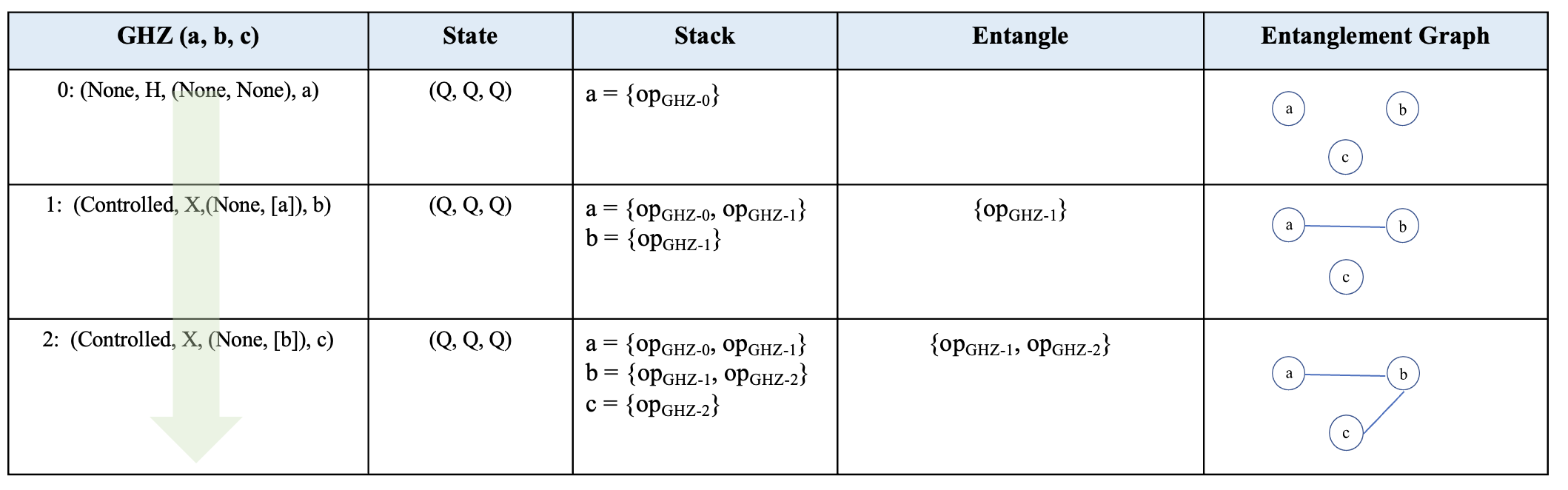}}
\centerline{\includegraphics[width=0.9\linewidth]{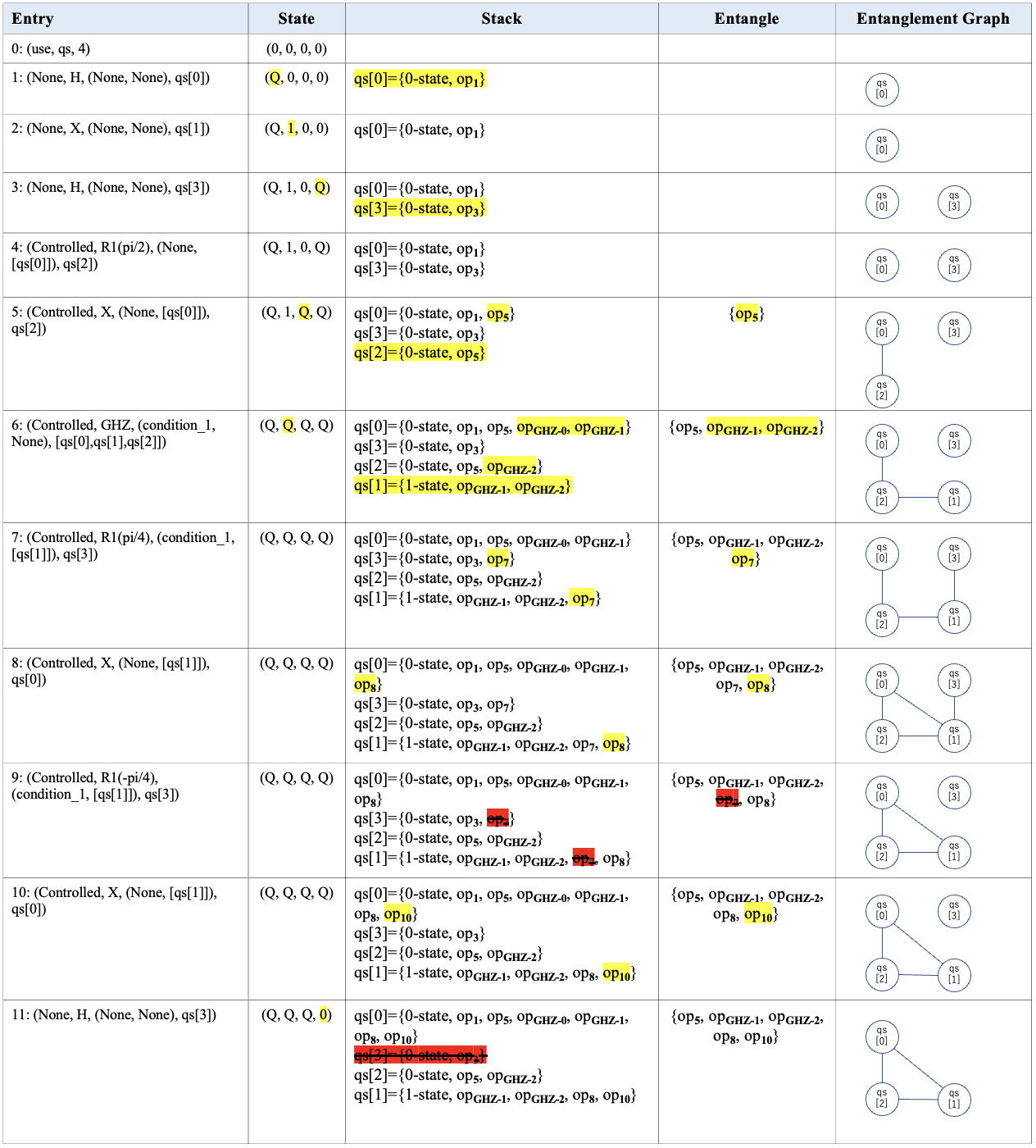}}
\caption{The intraprocedural analysis based on the CFGs for the \texttt{GHZ} operation and \texttt{Entangle\_test} operation.}
\label{fig:intraprocedural analysis}
\end{figure*}

For multiple calls to the \verb+GHZ+ function, we only need to perform the stack merge and graph transformation instead of repeatedly executing the \verb+GHZ+ function. The operations of line-7 and line-9 satisfy both the mutual inverse operation and the same control relation, so the execution of the line-9 operation will satisfy the uncomputation mechanism, so the line-7 operation will be canceled, and the corresponding entanglement relation will be eliminated. Finally, when the execution reaches the line-11 operation, the uncomputation mechanism is also satisfied, so the line-3 operation in \texttt{qs[3]} is also canceled. At this time, the processing of \texttt{qs[3]} by transition rule  $\mathcal{Q}$ will change to the initial 0-state. At this time, we will delete the corresponding stack and node. The final entanglement relation returned by our algorithm is shown in Figure~\ref{fig:intraprocedural analysis}.

\section{The Analysis Algorithm}
\label{sec:entanglement analysis}

We next present our entanglement analysis algorithm for Q\# programs. For each program point, the algorithm computes an entanglement graph for the current analysis scope at that point. The scope of the intraprocedural analysis is the currently analyzed \texttt{operation} module up to that point. The interprocedural analysis extends the scope to include the (transitively) called \texttt{operation} modules.
%

\subsection{Program Representation}
\label{subsec:program-representation}
We present several program representations which will be used by the algorithm during the analysis.

\vspace*{1.5mm}
\noindent
\textcolor{black}{
\textit{$\bullet$ Quantum Operation Notation.}
To better extract the information, we transform the quantum operation into the following operation line form:
\vspace*{0mm}
\begin{center}
\texttt{(Functor, Operation, Control, Target)}
\end{center}
\vspace*{0mm}
where \textit{Functor} represents the properties of the operation. There are four properties in Q\#: \textit{None, Adjoint, Controlled} and \textit{Adjoint Controlled}, which correspond to the original operation, the inverse operation, the controlled operation, and the inverse controlled operation, respectively. \textit{Operation} represents the name of the quantum operation. \textit{Control} represents the control relationship of the operation during execution, either as a classical \textit{condition} or as qubits control (\textit{Q-control}). Since these two types of control do not affect each other, in \textit{Control} we will use the expression (\textit{condition}, \textit{Q-control}). For the \textit{condition}, we use 0, 1, and unknown. In addition, to improve the accuracy of the analysis, we will also determine the equivalence between different conditions. For the \textit{Q-control}, we will describe it in detail in the next section. \textit{Target} represents the object of quantum operation execution.
}

\vspace*{1mm}
\noindent
\textit{$\bullet$ Interprocedural Control Flow Graph.}
The algorithm uses an interprocedural control flow graph (ICFG) to represent the computation of an entire Q\# program, which is a combination of the call graph and the control flow graphs (CFGs) of the program. Each CFG represents the program's computation of an \texttt{operation}. In the CFG, each node represents a statement, and each edge represents the control flow from one statement to another. 

The ICFG can be constructed in the following three steps.
First, the CFG for each \texttt{operation} in a Q\# program is constructed. We can use some traditional CFG construction algorithms such as~\cite{ferrante1987program}.
Second, the call graph of the program is constructed. We can use some sophisticated call graph construction algorithms such as~\cite{ryder1979constructing}.
Finally, we replace each node in the call graph with its corresponding CFG to form the ICFG of the entire program. 

\textcolor{black}{In Q\#, there exist 11 types of statements, which are \textit{call statement}, \textit{return statement}, \textit{fail statement}, \textit{variable declaration}, \textit{variable reassignment}, \textit{iteration, while statement}, \textit{repeat statement}, \textit{if statement}, \textit{conjugation}, and \textit{qubit allocation}. Since the Q\# language is targeted at classical-quantum mixed programs, some of these statements are only used for classical programs.} The algorithm analyzes only those statements that affect the entanglement information. We assume that the program has been pre-processed so that all statements relevant to the analysis are in one of the following forms:
\vspace*{0mm}
\textcolor{black}{
\begin{itemize}
\item \textbf{Call statement}: An operation call site of the form $l = op(l_{0},\ldots,l_{k})$. Each operation call site corresponds to an operation call site $m\in M$.
\item \textbf{Qubit allocation}: Quantum operations in Q\#.
\item \textbf{Iteration}: \textit{for variable in iterable\_object\{op\}}
\item \textbf{If statement}: \textit{if condition then \{$op_{1}$\} else \{$op_{2}$\}}
\item \textbf{Conjugation}: \textit{within \{$op_{1}$\} apply \{$op_{2}$\}}
\item \textbf{Repeat statement}: \textit{repeat \{$op_{1}$\} until condition (fixup\{$op_{2}$\})}
\end{itemize}
}

We can convert the above statements into CFG using our notations as follows:

\begin{itemize}
    \item \textbf{Call statement}: \textit{(Functor, Operation, Control, Target)}
    \item \textbf{Qubit allocation}: \textit{(None, Operation, None, Target)}
    \item \textbf{Iteration}: Since the loop length is known as soon as the iteration value is known. We can use \textit{for variable in iterable\_object\{(Functor, Operation, Control, Target)\}} to generate the CFG.
    \item \textbf{If statement}: \textit{(Controlled, $op_{1}$, condition, Target)} + \textit{(Adjoint Controlled, $op_{2}$, condition, Target)} + \textit{(None, $op_{2}$, None, Target)}
    \item \textbf{Conjugation}: This is an automated generation of uncomputation statements provided by Q\#. When \textit{$op_2$} in \textit{apply} block is executed, the inverse operation of \textit{$op_1$} in \textit{within} block is automatically generated. We convert the Conjugation statement into the following form: \textit{(None, $op_{1}$, None, Target)} + \textit{(None, $op_{2}$, None, Target)} + \textit{(Adjoint, $op_{1}$, None, Target)}
    \item \textbf{Repeat statement}: The repeat statement will execute the $op_{1}$ before evaluating the condition. If the condition is true, the loop exits. Otherwise, an additional $op_{2}$ defined in \textit{fixup} block (if present) will execute before entering the next loop iteration. We can convert the repeat statement to the following form: \textit{(None, $op_{1}$, None, Target)} + \textit{n * (Controlled, $op_{2}$, not-condition, Target)} + \textit{(Controlled, $op_{1}$, not-condition, Target)}] ($n$ is the maximum number of iteration)
\end{itemize}

After such a program pre-processing, only notations related to quantum operations are kept in the CFG.
The control flow graph for each \texttt{operation} $op$ starts with the \texttt{enter} statement $enter_{op}$ and ends with an exit statement $exit_{op}$.

The analysis represents the control flow relationships between statements as follows: $pred(st)$ is the set of statements that may execute immediately before $st$, and $succ(st)$ is the set of statements that may execute immediately after $st$. There are two program points for each statement $st$, the program point $\bullet$$st$ immediately before $st$ executes, and the program point $st$$\bullet$ immediately after $st$ executes.

The interprocedural analysis uses call graph information to compute sets of \texttt{operations} that may be called at \texttt{operation} call sites. For each \texttt{operation} call site $m$, $callees(m)$ is the set of \texttt{operations} that $m$ may call. Given an \texttt{operation} $op$, $caller(op)$ is the set of \texttt{operation} call sites that may call $op$. The current implementation obtains this call graph information using a variant of class hierarchy analysis. Still, the algorithm can use any conservative approximation to the actual call graph generated when the program runs.

\vspace*{1.5mm}
\noindent
\textit{$\bullet$ Entanglement Graph.}
The algorithm uses an entanglement graph to represent the entanglement relations between qubits at each program point of the analyzed program. In an entanglement graph, each node represents a qubit in a superposition state, and each edge represents the entanglement relation between two qubits. Two nodes in an entanglement graph are entangled if they are connected. The algorithm modifies the entanglement graph at each program point according to the ICFG of the program.

\subsection{Quantum State System}
\label{subsec:quantum-state}

\textcolor{black}{In quantum computing, the qubit state can only be $\ket{0}$, $\ket{1}$ or superposition. Qubits in the \textit{Q-control} will affect the execution result of the controlled operation when they are in different states. When in the \textit{Q-control} exists a qubit in $\ket{0}$, the operation will not be executed. For the qubit in the \textit{Q-control} that is in $\ket{1}$, even if the operation is executed, it cannot be entangled with the \textit{Target} qubit. Only the qubits in a superposition state can generate entanglement relations. Therefore, we construct a quantum state system: 0, 1, and $\mathcal{Q}$ correspond to $\ket{0}$, $\ket{1}$, or superposition, respectively. Only when the qubit is the execution target of an operation the state will be transitioned. Therefore we will analyze the state transition when each qubit is the target of the operation. We will describe the transition rules between these states as follows:
}

\vspace*{1.5mm}
\noindent
\textit{$\bullet$ \textcolor{black}{Transition rule 0:}}

\begin{table}[htbp]
\begin{center}
\scriptsize
\renewcommand\arraystretch{1}
\begin{tabular} {p{1.4cm}|p{5cm}}
\hline 
\textbf{Transition } & \textbf{Condition
} \\\hline 
\hline
$0 \rightarrow 0$: & The executed operations are non-magnitude operations. \\\hline 
$0 \rightarrow 1$: & The executed operations are NOT(X) operations which satisfy non-Q-Control or all qubits in Q-control are in 1-state.\\\hline
$0 \rightarrow \mathcal{Q}$: & The executed operations are HAD operations, or NOT(X) operations which satisfy in Q-control exists a qubit in $\mathcal{Q}$-state.
\end{tabular}
\end{center}
\end{table}

\noindent
\textit{$\bullet$ \textcolor{black}{Transition rule 1:}}

\begin{table}[htbp]
\scriptsize{
\begin{center}
\renewcommand\arraystretch{1}
\begin{tabular} {p{1.4cm}|p{5cm}}
\hline 
\textbf{Transition } & \textbf{Condition
} \\\hline 
\hline
$1 \rightarrow 0$: & The executed operations are NOT(X) operations which satisfy non-Q-Control or all qubits in Q-control are in 1-state under condition=1. \\\hline 
$1 \rightarrow 1$: & The executed operations are non-magnitude operations or NOT(X) operations which satisfy non-Q-Control, or all qubits in Q-control are in 1-state under condition=unknown.\\\hline
$1 \rightarrow \mathcal{Q}$: & The executed operations are HAD operations, or NOT(X) operations which satisfy in Q-control exists a qubit in $\mathcal{Q}$-state.
\end{tabular}
\end{center}
}
\end{table}

\textcolor{black}{A qubit in a superposition state will exist in an infinite number of states. If each state is recorded precisely, the computational cost of the program analysis will rise exponentially. Therefore, we use the stack structure to record the operation process of each $\mathcal{Q}$-state qubit. When 0-state and 1-state qubits are transitioned to $\mathcal{Q}$-states, we create a stack to record the quantum operations. The transition rule for $\mathcal{Q}$-states will also depend on the stack.}

\noindent
\textit{$\bullet$ \textcolor{black}{Transition rule $\mathcal{Q}$:}}

\begin{table}[htbp]
\scriptsize{
\begin{center}
\renewcommand\arraystretch{1}
\begin{tabular} {p{1.4cm}|p{5cm}}
\hline 
\textbf{Transition } & \textbf{Condition
} \\\hline 
\hline
$\mathcal{Q} \rightarrow 0/1$: & Due to the uncomputation mechanism, the state of the qubit is transitioned to the original 0-state or 1-state when the execution of the operation makes the stack of the Target qubit empty. \\\hline 
$\mathcal{Q} \rightarrow \mathcal{Q}$: & In other cases, the qubit will maintain the original $\mathcal{Q}$-state.
\end{tabular}
\end{center}
}
\end{table}

\subsection{Intraprocedural Analysis}
\label{subsec:intraprocedural-analysis}

\textcolor{black}{After pre-processing the Q\# program, we use the processed list that stores the operation lines as traversal objects and input them into the intraprocedural analysis. In our program analysis, we will introduce three parts to record the analysis results: (\textit{State}, \textit{Stack}, \textit{Entangle}). \textit{State} is used to record the state system of each qubit. \textit{Stack} is used to record the operation content of the qubit in $\mathcal{Q}$-state.  Due to the strict requirements of the uncomputation mechanism for the order of operations, we use the stack to record operations. \textit{Entangle} is used to record the entanglement relationship between qubits.}

\textcolor{black}{
For each operation line in CFG, we will denote the changes in the system by (\textit{State}, \textit{Stack}, \textit{Entangle}). First, we modify the \textit{State} using the transition rules we have defined. Then for the \textit{Stack}, we define the \textit{killing rule} $Kill_S$ and \textit{generation rule} $Gen_S$ as follows:
}

\begin{footnotesize}
$$Kill_{S}(x,op_i) = \begin{cases}
Stack_{IN}(x)[-1] & if\ x\ is\ op_i.Target and\ Stack_{IN}(x)[-1] \\
 & is\ inverse\ with\ op_i.\\
Stack_{IN}(op_i.Target)[-1] & if\ x\ is\ in\ op_i.Control\ and \\ 
& Kill_{S}(op_i.Target,op_i)\ne \emptyset\\
\emptyset & otherwise
\end{cases}$$
\end{footnotesize}

\begin{footnotesize}
$$Gen_{S}(x,op_i) = \begin{cases}
op_i & if\ (x\ is\ op_i.Target\ or\ x\ is\ in\ op_i.Control)\ and\\ 
& Kill_{S}(x,op_i)=\emptyset\\
\emptyset & otherwise
\end{cases}$$
\end{footnotesize}

\textcolor{black}{
In the $Kill_S$, we determine whether the new operation satisfies the uncomputation mechanism. If it does, we delete all the corresponding operation records \textit{$op_i$}. Otherwise, we add the operation to the \textit{Stack}. The transfer function for \textit{Stack} is as follows:}

\begin{footnotesize}
\begin{align*}
Stack_{OUT}(x, op_i) = (Stack_{IN} - Kill_{S}(x,op_i)) \cup  Gen_{S}(x,op_i)
\end{align*}
\end{footnotesize}

\textcolor{black}{Also for \textit{Entangle}, we define the corresponding \textit{killing rule} $Kill_E$ and \textit{generation rule} $Gen_E$ as follows:}

\begin{footnotesize}
$$Kill_{E}(op_i) = \begin{cases}
Stack_{IN}(op_i.Target)[-1] & if\ Kill_{S}(op_i.Target,op_i) \ne \emptyset \\ 
& and\ Stack_{IN}(op_i.Target)[-1]\\
& is\ in\ Entangle_{IN}\\
\emptyset & otherwise
\end{cases}$$
\end{footnotesize}

\begin{footnotesize}
$$Gen_{E}(x,op_i) = \begin{cases}
op_i & if\ Kill_{E} = \emptyset\ and\ \mathcal{Q}-state\ is\ in\ State[op_i.Control]\\
\emptyset & otherwise
\end{cases}$$
\end{footnotesize}

\textcolor{black}{The transfer function for \textit{Entangle} is as follows:}

\begin{footnotesize}
\begin{align*}
Entangle_{OUT}(op_i) = (Entangle_{IN} - Kill_{E}(op_i))\cup  Gen_{E}(op_i)
\end{align*}
\end{footnotesize}

\textcolor{black}{For the overall algorithm process, the first thing that needs to be verified is whether each operation is executed effectively in the current qubit state system, so we introduce a \textit{check\_executed} function. This function checks the \textit{Control} of each operation. When the \textit{condition} in the \textit{Control} exists as False or in the \textit{Q-control} exists a qubit in 0-state, it means that the operation will not be executed and can therefore be ignored.}

After ensuring that each operation can be executed efficiently, we will introduce the \textit{check\_fundamental} function to check the call relationship of the operation. When a call relationship is checked for an operation, we will use interprocedural analysis to process it. For the fundamental operations, we will continue to execute intraprocedural analysis and create an extensible library of fundamental operations for quick checking.

\begin{figure}[!h]
\centerline{\includegraphics[width=0.8\linewidth]{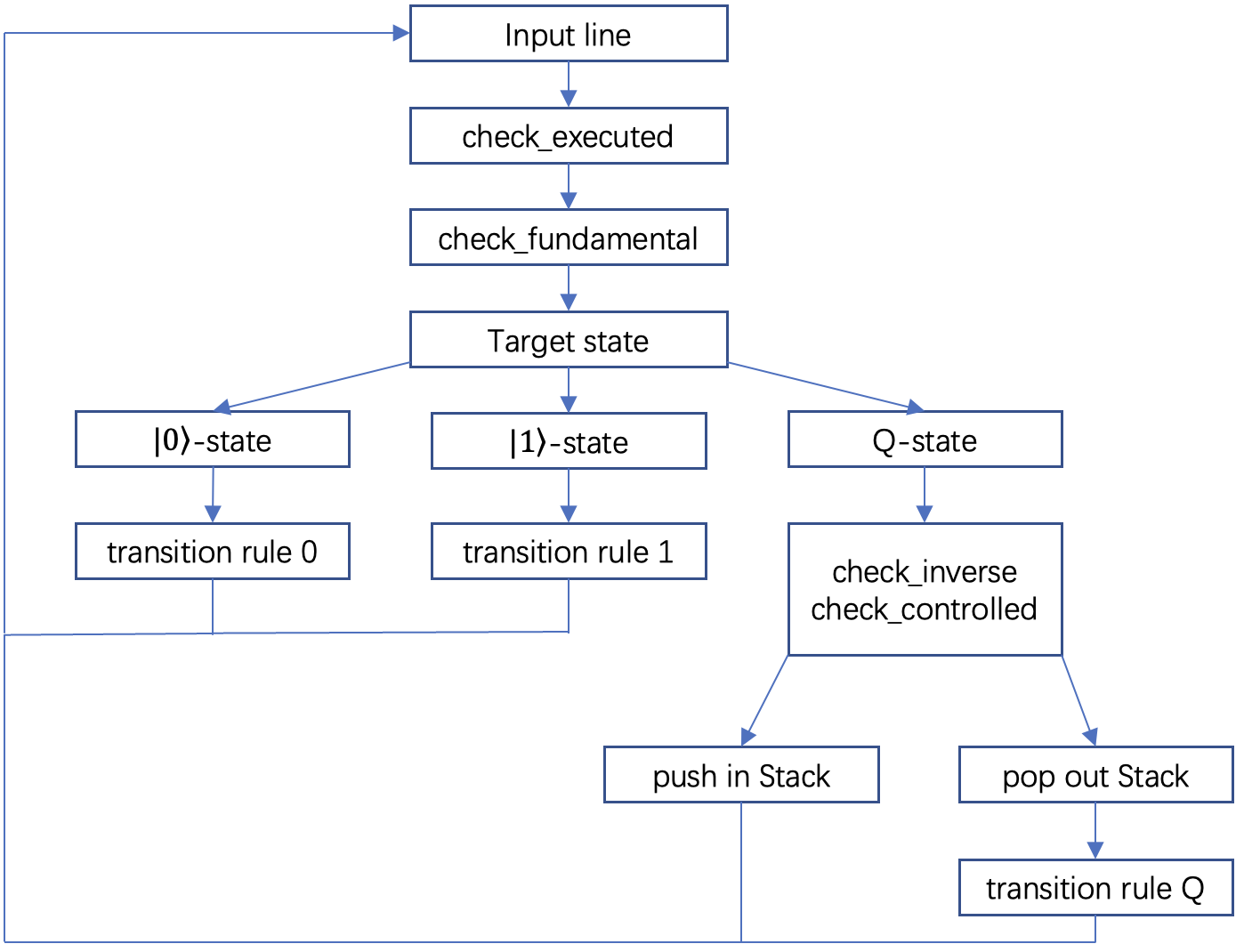}}
\caption{Algorithm flow of intraprocedural analysis}
\label{fig:intraprocedural flow}
\vspace{-2mm}
\end{figure}

Once it is ensured that the operation can be executed correctly and is in the processing range of the intraprocedural analysis, we process it differently depending on the state the Target qubit is in: For a qubit in 0-state or 1-state, we will use the transition rule 0 and transition rule 1, respectively, and modify the corresponding qubit state in \textit{State}. When a qubit exists that is transformed to $\mathcal{Q}$-state, we will create the corresponding operation stack in \textit{Stack} and push the original state of the qubit and the corresponding operation lines into the stack. For the $\mathcal{Q}$-state qubit, we check whether the current operation and the last operation in the corresponding stack satisfy the uncomputation mechanism. we introduce the \textit{check\_inverse} and \textit{check\_controlled} functions, which correspond to checking whether the two operations are mutually inverse and checking whether the control relationship between the two operations is the same, respectively. If both functions are satisfied, the current operation satisfies the uncomputation mechanism, so we pop out the operation in the corresponding stack and perform transition rule $\mathcal{Q}$ to detect the state transition of the target qubit. Since the current operation makes the original operation invalid, the entanglement relation generated by the original operation will be eliminated. We will remove the corresponding operation in the \textit{Entangle} and remove the corresponding edge in the entanglement graph. If the two functions are not satisfied at the same time, the current operation cannot eliminate the original operation, so we push the current operation into the stack. If there is a qubit in $\mathcal{Q}$-state in the \textit{Q-Control} of the current operation, then the current operation will create a new entanglement relationship. Therefore, we save the current operation in the \textit{Entangle} and create an edge of all $\mathcal{Q}$-state qubits in the \textit{Q-control} connected to the target qubit. The overall flow is shown in Figure~\ref{fig:intraprocedural flow}.

We will repeat this process until the operation list is fully traversed. Since existing quantum programs do not have dead loops, our method will have a stopping point. At this point, after the execution of the quantum program, the states of all qubits can be obtained from the State, and the entanglement relations in the program can be obtained from the entanglement graph.

\subsection{Interprocedural Analysis}
\label{subsec:interprocedural-analysis}

The algorithm uses a dataflow analysis to generate an entanglement at each point of the \texttt{operation}. The analysis of each \texttt{operation} starts with the construction of the entanglement graph (a, b, c) for the first statement in the \texttt{operation}. 

We perform an interprocedural analysis to deal with the problem of calls between \texttt{operation} modules. There are many function-specific modular functions in existing quantum programs, such as GHZ, QFT, and Amplitude Amplification (AA). We generate the corresponding stack and entanglement graphs for modular functions at high frequencies in advance. Unlike regular processing, the $\mathcal{Q}$-state is sensitive to all quantum operations since the call to the function requires input. In the face of quantum operations, the $\mathcal{Q}$-state behaves differently from the 0-state and 1-state. Therefore, we presuppose that all qubits of inputs are in the $\mathcal{Q}$-state when processing the calling \texttt{operation} module. When the main program executes the calling \texttt{operation} module, we create an alias relationship between the qubit as input in the main module and the qubit in the pre-processing calling \texttt{operation} module. Then the two stacks are merged, and the stack in the calling module is stacked from the bottom of the stack to the stack in the main module in order. If the inverse operation is satisfied, the offset is performed; otherwise, the stack is entered. For the entanglement graph, the qubit of the main operation inherits the concatenation relation in the calling operation and then removes the nodes in the calling operation graph.

If the qubit of the main operation cannot maintain the $\mathcal{Q}$-state after the merging of two stacks is finished, the concatenation relation is passed, and the node is deleted.

\textcolor{black}{Since Q\# can generate various versions of operation (e.g., \textit{Controlled}) when calling \texttt{operation}, depending on the version used, we also need to adjust the corresponding operation line. Therefore, we will propose a \textit{line wrapper} that adjusts the operation line inside the called operation according to the properties of the calling \texttt{operation}, modifying the corresponding \textit{Functor} and \textit{Control}.}



%



\subsection{Soundness}
\label{subsec:soundness}
\textcolor{black}{In the existing method, all statements that may generate entanglement are considered entanglement relations. Although this method can guarantee the soundness of the entanglement relations, there will be a large number of false positive entanglement relations. In our approach, the entanglement relation in the program is not monotonically increasing due to the introduction of the uncomputation mechanism. Therefore, concerning the soundness of our method, it is sufficient to argue that the entanglement relations eliminated after the introduction of the uncomputation mechanism are all false positives.}

\textcolor{black}{The uncomputation mechanism relies on two parts of the operations: whether the two operations are inverse operations and whether the control relationship between the two operations is consistent. Our approach does not introduce equivalent operations at this stage for the first case of inverse operations, so we define inverse operations conservatively using basic knowledge of quantum computing and \textit{Adjoint} functor in Q\#. For the second case of control relations, we divide it into two kinds of control relations: quantum control and classical control. For quantum control, we will determine whether the two operations are controlled by the same qubits and will determine whether the control qubits are in the same magnitude state. For classical control, we will compare whether the classical constraints of the two operations are exactly the same.}

\textcolor{black}{The determination of the uncomputation mechanism by our method allows us to eliminate entanglement relations only for absolute inverse operations. Therefore, the entanglement relations we eliminate belong to the false positive relations that should be uncomputed in any case. The entanglement relations that can be eliminated only in some real cases will not be considered satisfying the uncomputation mechanism, so some of the false positive relations will be retained in our method.}

\textcolor{black}{In addition, since the entanglement relation only arises between superposition qubits and the invalid execution of some operations, the state system we define also eliminates some false positive relations.}

\section{Proposed Implementation}
\label{sec:implementation}

\begin{figure}[!h]
\centerline{\includegraphics[width=0.9\linewidth]{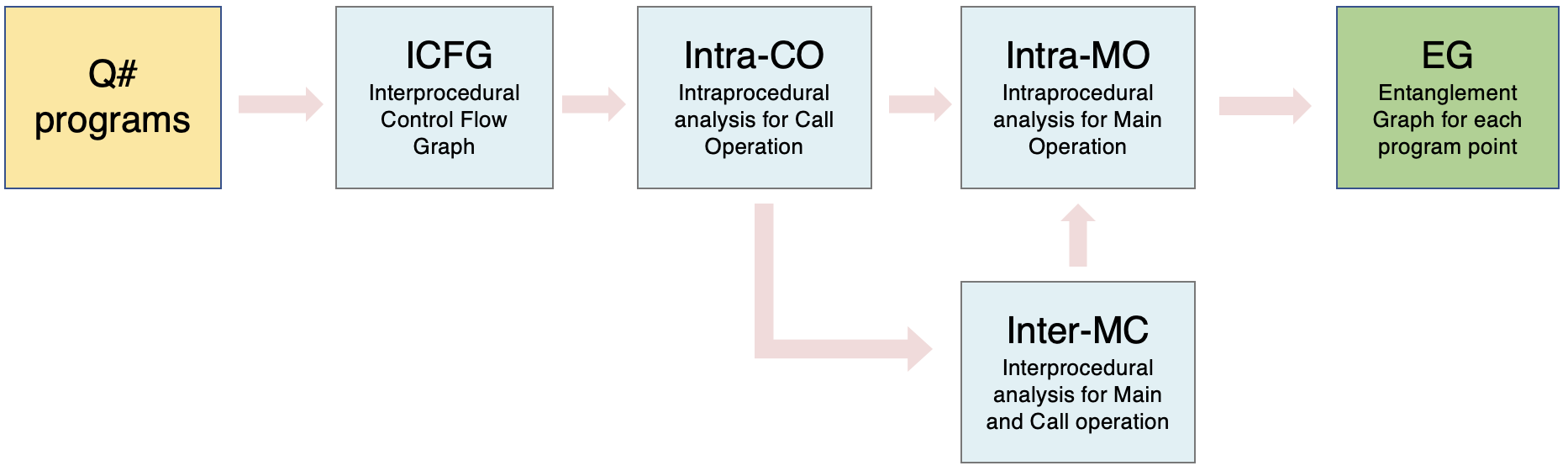}}
\caption{A framework of \textit{JiuChan} tool}
\label{fig:jiuchan}
\end{figure}

We plan to develop a tool called \textit{JiuChan} to support our method. Figure~\ref{fig:jiuchan} shows the framework of \textit{JiuChan}: the inputs are the Q\# program, the call graph of the program, which is generated by using the algorithm in~\cite{ryder1979constructing}, and the CFG of each operation which is generated by using the algorithm~\cite{ferrante1987program}; the outputs are the entanglement graphs (EGs) at all program points. For a Q\# program, we first generate the corresponding ICFG and transform each quantum operation into the structural form. Then, for all call operations, we first generate the corresponding Stack, Entangle, and Entanglement graphs using intraprocedural analysis. Then we perform intraprocedural analysis on the main operation, and when we encounter a call operation during the analysis, we perform interprocedural analysis using the already analyzed results. The relationship of each program node is generated and expressed in an undirected graph.

\label{sec:uses}

\section{Related Work}
\label{sec:related-work}
This section discusses some related work in the entanglement analysis of quantum programs.

As a first step in dealing with entanglement analysis, Perdrix~\cite{perdrix2007quantum} proposed a type system reflecting the entanglement and separability between quantum bits that can approximate the entanglement relations of an array of quantum bits. Prost and Zerrari~\cite{prost2009reasoning} proposed a logical entanglement analysis method to deal with functional programming languages with higher-order functions. They followed the idea of classical aliasing analysis proposed by Berger~\textit{et al.}~\cite{berger2005logical} and applied it to quantum entanglement analysis. Their logical framework can analyze more complex quantum programs, but not quantum programs without annotations and considering only pure quantum states. In \cite{perdrix2008quantum}, Perdrix further proposed an approach to entanglement analysis based on abstract interpretation~\cite{cousot1977abstract}. This approach establishes a correlation between concrete quantum semantics and a simple quantum programming language based on super operators.  ScaffCC~\cite{javadiabhari2015scaffcc} supports conservative entanglement analysis by identifying each pair of quantum bits that may be entangled together in a program. The resulting entanglement information can help programmers design algorithms and perform debugging. ScaffCC uses data flow analysis techniques to obtain entanglement information in programs. Yuan~\textit{et al.}~\cite{yuan2022twist} formalized purity as a central tool for automatically reasoning about entanglement problems in quantum programs. A pure expression is one whose evaluation is not affected by measurements of qubits it does not own, meaning no entanglement with any other expression in the computation. They also designed \textit{Twist}, the first language with a type system for reasoning about purity.
However, these methods only analyze fine-grained entanglement between specific qubits, and the scale of their analysis is limited. In addition, they only consider simple quantum programming languages but cannot analyze practical ones. In contrast, to the best of our knowledge, \textit{JiuChan} is the first static entanglement analysis method aimed at analyzing entanglement relations in a practical quantum programming language Q\# for handling large-scale quantum programs.

\section{Conclusion}
\label{sec:conclusion}

This paper has presented a static entanglement analysis method for quantum programs developed in the practical quantum programming language Q\#. Our method first constructs an interprocedural control flow graph (ICFG) for a Q\# program to perform the analysis. It then calculates the entanglement information within each module and between program modules. Our analysis approach can help improve the reliability and security of quantum programs by uncovering entanglement-induced errors in the programs.

As for future work, we would like to handle more language features in Q\#, such as classical-quantum mixed programs, and apply our analysis approach to other quantum programming languages.
We are implementing an entanglement analysis tool for Q\# to demonstrate its effectiveness. The next step for us is to perform some experiments to evaluate the usefulness of our entanglement analysis in practical quantum software development. 

\bibliography{qse-bibliography}
\bibliographystyle{plain}


\end{document}